%% file: main.tex
\documentclass{edm_article}

\input{macros}
\begin{document}

\title{Humanizing Automated Programming Feedback: Fine-Tuning Generative Models with \\ Student-Written Feedback}

\numberofauthors{7}
\author{
\alignauthor
Victor-Alexandru P{\u a}durean\\
       \affaddr{MPI-SWS}\\
       \email{vpadurea@mpi-sws.org}
\alignauthor
Tung Phung\\
       \affaddr{MPI-SWS}\\
       \email{mphung@mpi-sws.org}
\alignauthor
Nachiket Kotalwar\\
       \affaddr{MPI-SWS}\\
       \email{nkotalwa@mpi-sws.org}
\and
Michael Liut\\
       \affaddr{University of Toronto Mississauga}\\
       \email{michael.liut@utoronto.ca}
\alignauthor
Juho Leinonen\\
       \affaddr{Aalto University}\\
       \email{juho.2.leinonen@aalto.fi}
\alignauthor
Paul Denny\\
       \affaddr{University of Auckland}\\
       \email{paul@cs.auckland.ac.nz}
\and
Adish Singla\\
       \affaddr{MPI-SWS}\\
       \email{adishs@mpi-sws.org}
}

\maketitle

\input{0_abstract}


\input{1_introduction}

\input{2_relatedwork}

\input{3_setup}

\input{4_results}

\input{5_conclusion}

\input{6_acknowledgements}

\bibliographystyle{abbrv}

\bibliography{main}

\end{document}

%% file: macros.tex
\usepackage{etoolbox}
\usepackage{booktabs} 
\usepackage{graphicx}
\usepackage{wrapfig}
\usepackage{multirow}
\usepackage{varwidth}
\usepackage{subcaption}

\usepackage{enumitem}
\usepackage{xcolor}
\usepackage{pifont}
\usepackage{colortbl}
\usepackage[numbers,sort&compress]{natbib}
\usepackage{url}

\usepackage{soul}
\sethlcolor{yellow!25}

\definecolor{RoyalPurple}{RGB}{120, 81, 169}
\definecolor{OliveGreen}{RGB}{128, 128, 0}

%% file: 0_abstract.tex
\begin{abstract}
The growing need for automated and personalized feedback in programming education has led to recent interest in leveraging generative AI for feedback generation. However, current approaches tend to rely on prompt engineering techniques in which predefined prompts guide the AI to generate feedback. This can result in rigid and constrained responses that fail to accommodate the diverse needs of students and do not reflect the style
of human-written feedback from tutors or peers. In this study, we explore learnersourcing as a means to fine-tune language models for generating feedback that is more similar to that written by humans, particularly peer students. Specifically, we asked students to act in the flipped role of a tutor and write feedback on programs containing bugs. We collected approximately $1,900$ instances of student-written feedback on multiple programming problems and buggy programs. To establish a baseline for comparison, we analyzed a sample of $300$ instances based on correctness, length, and how the bugs are described. Using this data, we fine-tuned open-access generative models, specifically Llama3 and Phi3. Our findings indicate that fine-tuning models on learnersourced data not only produces feedback that better matches the style of feedback written by students, but also improves accuracy compared to feedback generated through prompt engineering alone, even though some student-written feedback is incorrect. This surprising finding highlights the potential of student-centered fine-tuning to improve automated feedback systems in programming education.
\end{abstract}

%% file: 1_introduction.tex

\section{Introduction}\label{sec.introduction}

Generative AI offers great potential to enhance programming education by providing personalized feedback to students. This can complement feedback from human tutors by offering continuous, around-the-clock support~\cite{DBLP:conf/sigcse/WangMP24,DBLP:journals/corr/abs-2402-01580}. However, students often perceive AI-generated feedback as less effective than human-written feedback because it may not fully address their needs or provide the depth of support they expect~\cite{otaki2024generative}.  In particular, students value the emotional sensitivity of human feedback. Therefore, to maximize student adoption, it is essential to refine AI models to deliver feedback with human-like characteristics, similar to what students would receive from tutors or peers.

\input{figs/finetuning/examples_llama_fig_overview}

Most previous studies have focused on the use of prompt-engineering techniques to customize feedback to meet expert educators' standards \cite{qi2023conversational,DBLP:journals/corr/abs-2310-13712}. Furthermore, the quality of feedback is typically evaluated via educator-created rubrics \cite{DBLP:journals/corr/abs-2306-17156,DBLP:conf/lak/PhungPS0CGSS24}, emphasizing conciseness and appropriateness (i.e., not explicitly pointing out how to fix bugs). This approach suffers from several limitations. First, it demands extensive effort from experts to engineer prompts which may not generalize across domains. Second, the generated feedback tends to be rigid and constrained (e.g., spanning one or two sentences and not containing code), lacking flexibility for addressing  specific needs and different types of bugs. More generally, this approach does not capture student preferences as well as the dynamic and adaptable characteristics of humans, such as tutors and peers.

To address this gap, we leverage learnersourced data for fine-tuning generative models, combining the relatability of student-written feedback with generative models' ability to generate consistent feedback at scale. By fine-tuning on $1,920$ student-created feedback instances, we align feedback generated by AI more closely with student communication styles, making it more concise, peer-like, and adaptable, as illustrated in Figure~\ref{fig.finetuning.examples_llama}. This approach reduces reliance on manual prompt engineering while enhancing the human-like qualities of automated feedback in programming education. Our contributions are as follows: 

\setlength{\leftmargini}{2em}
\begin{enumerate}[label=\Roman*.] 
\item We introduce a fine-tuning method using learnersourced data to align AI-generated feedback with student-written feedback, reducing reliance on prompt engineering.

\item We propose a flipped-role setup for feedback collection and define key attributes for analyzing feedback style.

\item Our fine-tuned models produce feedback that is both more similar to student-written feedback and more accurate than basic and prompt-engineered approaches.

\item We publicly release our fine-tuning implementation to support future research in educational data mining.

\end{enumerate}

%% file: figs/finetuning/examples_llama_fig_overview.tex
\begin{figure*}[t!]
\Description{Three side-by-side feedback examples from Llama3-8B with different prompts: default, engineered, and fine-tuned. The first is verbose, the second is concise but inaccurate, and the third is accurate and actionable.}
\centering
	\begin{minipage}{\linewidth}
    {
        \begin{subfigure}{0.99\linewidth}
        {\centering
            \setlength{\fboxsep}{2pt}\fbox{\begin{varwidth}{\dimexpr\textwidth-2\fboxsep-2\fboxrule\relax}
            \small
            \input{figs/finetuning/src/base_llama.tex}
            \end{varwidth}
            }
            \caption{Feedback with Llama3-8B and basic prompt.}		
            \hfill
            \label{fig.finetuning.examples_llama.base}
        }
        \end{subfigure}
    }
    \end{minipage}
    \begin{minipage}{\linewidth}
    {
        \begin{subfigure}{0.54\linewidth}
        {\centering
            \setlength{\fboxsep}{2pt}\fbox{\begin{varwidth}{\dimexpr\textwidth-2\fboxsep-2\fboxrule\relax}
            \small
            \input{figs/finetuning/src/prompting_llama.tex}
            \end{varwidth}
            }
            \caption{Feedback with Llama3-8B and engineered prompt.}				
            \label{fig.finetuning.examples_llama.prompting}
        }
        \end{subfigure}
        \ \ \ \
        \begin{subfigure}{0.44\linewidth}
        {\centering
            \setlength{\fboxsep}{6.5pt}\fbox{\begin{varwidth}{\dimexpr\textwidth-2\fboxsep-2\fboxrule\relax}
            \small
            \input{figs/finetuning/src/finetuned_llama.tex}
            \end{varwidth}
            }
            \caption{Feedback with Llama3-8B-fine-tuned and basic prompt.}				
            \label{fig.finetuning.examples_llama.finetuned}
        }
        \end{subfigure}
    }
    \end{minipage}
     \caption{Examples of feedback by Llama3-8B with different settings for a buggy program attempting the ``Sum Positive Values'' problem (see~Figure~\ref{fig.illustration}). Feedback with the fine-tuned model is correct in contrast to the feedback given with the engineered prompt which contains redundant information that may confuse the student.
     }
    \label{fig.finetuning.examples_llama}
\end{figure*}

%% file: figs/finetuning/src/base_llama.tex
I'd be happy to help!\newline
The main issue with your original program is that you're adding the index `i' to the sum instead of the value at that index `values[i]'. This is why you're getting the wrong output.\newline
Here's a breakdown of what's happening:\newline
* When `i' is 0, `values[i]' is 10, which is positive, so you add 0 to the sum (because `i' is 0).\newline
* When `i' is 1, `values[i]' is 20, which is positive, so you add 1 to the sum (because `i' is 1).\newline
* When `i' is 2, `values[i]' is 30, which is positive, so you add 2 to the sum (because `i' is 2).\newline
So, the sum becomes 0 + 1 + 2 = 3, which is why you're getting the output 3.\newline
By changing the line `sum += i;' to `sum += values[i];', you're correctly adding the positive values to the sum, which is why the fixed program produces the correct output.\newline
Here are some additional suggestions to improve your code:\newline
* Consider adding a check to ensure that the input array is not null before trying to access its elements. This can help prevent crashes or unexpected behavior if the input array is invalid.\newline
* You might want to consider adding a comment to explain what the function does and what it returns. This can make it easier for others (or yourself!) to understand the code.\newline
* If you're working with large arrays, you might want to consider using a more efficient algorithm or data structure to store and process the values. However, for small arrays like the one in your example, the simple loop-based approach should be sufficient.\newline
Overall, your code is simple and easy to understand, and the fix is a good one. With a few minor tweaks, it should be even better!

%% file: figs/finetuning/src/prompting_llama.tex
Think about what you're actually trying to sum up in your loop. Are you trying to sum up the indices of the array, or the values themselves? Take a closer look at the condition in your if statement and see if it's doing what you intend it to do.

%% file: figs/finetuning/src/finetuned_llama.tex
The problem with the code is where you are adding the sum. Rather than adding the value of the element, you are adding the index of that element.

%% file: 2_relatedwork.tex
\section{Related Work}\label{sec.relatedwork}
\looseness-1\textbf{Expert-designed feedback.}
Even prior to the advent of generative AI, significant efforts were made to provide automated feedback to students in introductory programming courses. Early work utilized unit test-based feedback in automated assessment systems, where students are shown specific error messages depending on which tests pass or fail~\cite{ihantola2010review,paiva2022automated,ala2005survey}. Other studies relied on expert-crafted or learned rules to identify specific bugs, thereby providing tailored feedback \cite{singh2013automated,kohn2020tell,piech2015learning,becker2016effective}. While these methods ensured high-quality feedback, they required substantial expert involvement and large training datasets, limiting scalability.

\textbf{Crowdsourced programming feedback.}
To overcome the limitations of expert-designed feedback, researchers explored crowdsourcing to collect errors and fixes contributed by the programming community~\cite{head2017writing,mujumdar2011crowdsourcing,jiang2018toward,al2018review}. Initiatives like HelpMeOut~\cite{hartmann2010would} focused on collecting and matching code fixes for buggy programs, but these methods faced challenges in generalizability and struggled to adapt to new bug types. Learnersourcing builds on this concept by engaging students directly in creating feedback~\cite{pirttinen2022can,singh2021s,singh2022learnersourcing}, which has been shown to generate valuable, relatable feedback and improve student learning~\cite{pirttinen2023lessons,lahza2023analytics}. Learnersourcing remains relevant alongside generative AI \cite{khosravi2023learnersourcing}, with recent studies exploring their synergy by examining student-written hints for programming bugs, both with and without AI assistance \cite{singh2024bridging}. In contrast, our work focuses on leveraging learnersourced data to enhance generative AI models for better feedback generation.

\textbf{Generative AI-powered programming feedback.}
The rise of generative models has opened new possibilities for automated feedback in education. Tools like GPT-3 and Codex~\cite{GPT-Family} have been used to enhance compiler error messages and provide syntax bug fixes~\cite{DBLP:conf/sigcse/0001HSRDPB23,DBLP:conf/edm/PhungCGKMSS23}. Efforts to improve the quality of AI-generated feedback include incorporating symbolic information~\cite{DBLP:conf/lak/PhungPS0CGSS24,qi2023conversational,islam2024mapcoder}, validation mechanisms~\cite{DBLP:conf/edm/PhungCGKMSS23,DBLP:conf/lak/PhungPS0CGSS24}, and retrieval-augmented generation (RAG)~\cite{DBLP:conf/sigcse/LiuZLHTM24,frazier2024customizing}. However, these studies have primarily focused on generating feedback that meets the standards of expert educators, often overlooking student preferences. Our work bridges this gap by focusing on generating feedback that aligns with learnersourced data, thereby enhancing the overall experience.

\input{figs/illustration/fig_overview}
\input{figs/gathering_data/gathering_data}

\textbf{Language model fine-tuning.}
Recent work has explored fine-tuning generative models, particularly small open-access models like Llama3-8B~\cite{Llama-3} and Phi3-3.8B~\cite{DBLP:journals/corr/abs-2404-14219}, to optimize for specific tasks in education~\cite{DBLP:journals/corr/abs-2403-13372,DBLP:conf/iticse/LiuYHBBL24}. Fine-tuning with synthetically generated data has been shown to significantly improve feedback quality for programming tasks~\cite{kotalwar2024hintsinbrowser}. In contrast, we use real student-written hints for fine-tuning, aiming to mimic human-like styles while improving the accuracy and alignment of AI-generated feedback.

%% file: figs/illustration/fig_overview.tex
\begin{figure*}[t!]
\Description{Boxes showing the problem description, a buggy program, the instruction for the flipped role task, and a genuine student-written feedback.}
            \begin{subfigure}[b]{0.46\linewidth}
            {
                \centering
                \begin{tabular}{|p|}
                    \hline
                    \multicolumn{1}{|c|}{\textbf{\color{RoyalPurple}Problem Description}} \\
                    \multicolumn{1}{|p{\linewidth}|}{
                        \input{figs/illustration/src/description_simple.tex}
                    }\\
                    \hline
                \end{tabular}
                \begin{tabular}{|p|}
                    \hline
                    \multicolumn{1}{|c|}{\textbf{\color{RoyalPurple}Buggy Program}} \\
                    \multicolumn{1}{|p{\linewidth}|}{
                        \includegraphics[width=\linewidth]{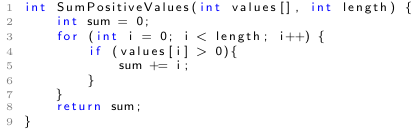}
                    }\\
                    \hline
                \end{tabular}
                \subcaption{Scenario}				
                \label{fig.illustration.scenario}
            }
            \end{subfigure}
        \ \ \ \ \
            \begin{subfigure}[b]{0.47\linewidth}
            {
            \centering
                \renewcommand{\arraystretch}{1.85}
                \begin{tabular}{|p|}
                    \hline
                    \multicolumn{1}{|c|}{\textbf{\color{RoyalPurple}\cellcolor{gray!20}
                    Instruction for the Flipped Role Task}} \\
                    \multicolumn{1}{|p{\linewidth}|}{
                        \input{figs/illustration/src/student_prompt.tex}
                    }\\
                    \multicolumn{1}{|c|}
                    {\textbf{\color{RoyalPurple}Student-Written Feedback}} \\
                    \multicolumn{1}{|p{\linewidth}|}{
                        \input{figs/illustration/src/feedback1.tex}
                    }\\
                    \hline
                \end{tabular}
                
                \subcaption{Student acts in a flipped role as a tutor}				
                \label{fig.illustration.student_prompt}
            }
            \end{subfigure}
     \caption{
    \textbf{(a)} shows the scenario including the problem description and the buggy code. \textbf{(b)} shows the instruction given to the student asking them to act in the flipped role as a tutor and give feedback and a genuine example of student-written feedback.
    }
    \label{fig.illustration}
\end{figure*}

%% file: figs/illustration/src/description_simple.tex
\textbf{Sum Positive Values}

Define a function called SumPositiveValues() which is passed two inputs: an array of integers, and an integer indicating how many elements are in the array. The function should return the sum of all positive integers in the input array.

%% file: figs/illustration/src/student_prompt.tex
\cellcolor{gray!20}
Imagine you are a tutor. A student in the course is asking for your help. Here is the problem description and the student's buggy C program. You should provide feedback so that the student can understand the issues in their buggy program and fix it. Provide your feedback in the textbox below.

%% file: figs/illustration/src/feedback1.tex
The program needs to output the sum of all the positive values.
This program is outputting the sum of all of the indexes of the values.
To fix it make sure you add the values not the index i.

%% file: figs/gathering_data/gathering_data.tex
\begin{figure}
\Description{A flowchart illustrating a flipped-role feedback task: students see buggy code and output, receive a fixed version, and are asked to write tutor-style feedback.}
    \centering
    {
        \includegraphics[width=0.98\linewidth]{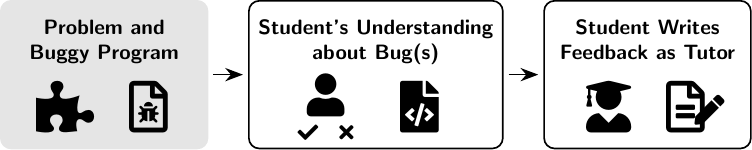}
        \caption{Feedback-writing task (Sections~\ref{sec.setup.task}~and~\ref{sec.setup.student})}
        \label{fig.introduction.gathering_data}  
    }
\end{figure}

%% file: 3_setup.tex
\section{Methodology}\label{sec.setup}

This section details our methodology for collecting and analyzing learnersourced feedback, as well as generating and evaluating AI-generated feedback. First, we describe the flipped-role feedback-writing task used to collect student-written feedback. Next, we outline the dataset, including problems, buggy programs, and collected responses. We then present our approach for feedback generation using different prompting strategies and our fine-tuning techniques. Finally, we explain the evaluation setup, including the rubric and expert annotation process used to analyze the attributes of the learnersourced and AI-generated feedback.

\subsection{Feedback-Writing Task} \label{sec.setup.task}

\textbf{Flipped role task.}
To leverage learnersourcing, we designed a feedback-writing task where students take on the role of tutors, providing feedback on buggy code. Figure~\ref{fig.introduction.gathering_data} illustrates the overall process, detailing the key components of the task. Students receive a problem description, a buggy implementation, and an instruction asking them to provide feedback from a tutor’s perspective. This setup encourages students to analyze errors, articulate their reasoning, and mimic the peer feedback process in programming education. Figure~\ref{fig.illustration} presents an example of this task, showing a specific problem along with the buggy code, the instruction given to the student, and a genuine example of student-written feedback. Since students typically have less programming experience than real tutors, providing meaningful feedback can be challenging. To support them, we incorporated a structured pre-feedback activity to help them better understand the bug before writing feedback.

\textbf{Pre-feedback activity.}
\looseness-1When presented with the problem and the buggy code, students first complete an activity where they are required to provide an input for a failing test case that would trigger a bug in the program, the buggy output on this input, and the correct output for this input. Based on whether they successfully complete this step, we categorize their understanding into two groups: \emph{Understanding=1} (successful) and \emph{Understanding=0} (unsuccessful). Regardless of their performance, all students are then shown a fixed version of the buggy code to help them better understand the problem and the bug. Finally, they write feedback on the buggy program from a tutor’s perspective, aiming to explain the issue and suggest how to fix it. This structured process ensures students reflect on errors before providing guidance.

\subsection{Student-Written Feedback Data} \label{sec.setup.student}
\textbf{Set of feedback-writing tasks.}
To prepare tasks, we started with $3$ programming problems and $10$ buggy programs for each. The problems and buggy programs were pre-selected by experts to cover different introductory programming concepts and capture diverse types of bugs. For each task, the student is given a problem and a buggy program and asked to act in the flipped role of a tutor to write feedback for that buggy program. In total, we obtained $30$ feedback-writing tasks. Below are the three problem descriptions:

\setlength{\leftmargini}{1em}
\begin{itemize}
\vspace{-2mm}
    \item Problem $1$: ``Sum Positive Values'' -- Return the sum of the positive values in the input array.
    \vspace{-0.5mm}
    \item Problem $2$: ``Print Summary'' -- Print ``Positive'', ``Negative'', or ``Equal'', depending on whether there are more positive or negative numbers in the input array until the terminating value $0$.
    \vspace{-0.5mm}
    \item Problem $3$: ``Print Average Rainfall'' -- Compute the average of non-negative integer values in the input array until the terminating value of $-999$.
\vspace{-2mm}
\end{itemize}

In terms of difficulty, Problem $1$ is generally considered simpler, while Problems $2$ and $3$ are more challenging.

\textbf{Course and students.}
The data collection process used a setup similar to a prior work \cite{DBLP:conf/sigcse/Padurean0S25}. Following this approach, we conducted our study in an introductory C programming course at the University of Auckland. There are about $750$ students enrolled in the course, who typically have little to no prior experience in programming. To allow for a large number of students to work on the feedback-writing tasks, we developed a web application and deployed it as part of a laboratory exercise, conducted towards the end of the course. During this exercise, each student was given three feedback-writing tasks, one for each problem in a fixed order (Problem 1 $\xrightarrow{}$ Problem 2 $\xrightarrow{}$ Problem 3).

\textbf{Data.}
Over $700$ students completed all three tasks. Regarding providing a correct failing test case that identified a bug, $503$ ($67.9$\%) students succeeded in Problem $1$, $308$ ($41.6$\%) in Problem $2$, and $314$ ($42.4$\%) in Problem $3$. After excluding empty entries, we collected $1920$ feedback instances across all tasks. The feedback ranged from concise single-word responses to detailed explanations of up to $295$ words. In general, the feedback consisted of complete well-structured sentences, as shown in Figure~\ref{fig.illustration.student_prompt} (bottom).

\subsection{AI-Generated Feedback Data} \label{sec.setup.technique}

\textbf{Technique for AI-generated feedback.} To investigate AI-generated feedback, we use a technique grounded in literature, which involves leveraging a problem description and a buggy code, along with additional symbolic information (failing test case and a fixed version of the code) to generate feedback \cite{DBLP:journals/corr/abs-2306-17156,DBLP:conf/lak/PhungPS0CGSS24,qi2023conversational}.

\input{figs/setup/study_workflow}

The fixed version of the code is typically obtained using a generative model with a separate prompt before asking the model for feedback.\footnote{As the focus of this study is to investigate AI-generated feedback in terms of style and structure, we used the same fixed code for all models when asking for feedback -- this allows us to compare more directly the impact of prompt instructions and fine-tuning.} Beyond this information, another important aspect is the instruction given to guide the model regarding the style of feedback it should generate. Here, we will investigate two prompting strategies:

\setlength{\leftmargini}{1em}
\begin{itemize}
\vspace{-3mm}
    \item Basic prompt: This prompt uses a basic instruction for the model to generate feedback without any style guidance (see Figure~\ref{fig.prompt_basic}).
    \vspace{-0.5mm}
    \item Engineered prompt: This prompt adopts instructions from techniques shown to be effective, developed through expert engineering in existing work \cite{DBLP:conf/lak/PhungPS0CGSS24,qi2023conversational} (see Figure~\ref{fig.prompt}).
\vspace{-4mm}
\end{itemize}

\textbf{Generative models.} Our analysis involves models from OpenAI's GPT family, specifically GPT-3.5 Turbo \cite{ChatGPT,GPT-Family} and GPT-4 Turbo \cite{GPT4,GPT-Family}, which have previously demonstrated their efficacy in providing feedback for debugging purposes \cite{DBLP:journals/corr/abs-2306-17156,DBLP:conf/lak/PhungPS0CGSS24}. Additionally, we used small open-access models for fine-tuning, namely Llama3-8B \cite{Llama-3} and Phi3-3.8B \cite{DBLP:journals/corr/abs-2404-14219} due to their growing popularity in educational settings for a lower cost, convenience, and better data privacy \cite{kotalwar2024hintsinbrowser,koutcheme2024open}. 

\textbf{Fine-tuning setup.}
To examine the effects of model fine-tuning on generated feedback, we conducted supervised fine-tuning on Llama3-8B and Phi3-3.8B with student-written feedback\footnote{The code used for fine-tuning these models is available at \url{https://github.com/machine-teaching-group/edm2025-humanizing-feedback}.}. From the total of $1920$ collected instances, we applied a filtering step to keep only those between $5$ and $200$ words, resulting in $1903$ feedback instances. Our fine-tuning prompt mirrors the basic prompt as shown in Figure~\ref{fig.prompt_basic} but includes formatting instructions (i.e., asking the model to include in its answer a start and an end token). We picked hyperparameters based on existing literature~\cite{DBLP:conf/nips/ZhouLX0SMMEYYZG23}.

\subsection{Evaluation Setup}\label{sec.setup.rubric}
\textbf{Evaluation rubric.}
Our aim is to first understand student-written feedback characteristics and then align generative models with them. To achieve this, we develop a detailed rubric capturing various attributes of feedback:

\input{figs/finetuning/prompt_overview}
\input{figs/results/fig_overview}

\setlength{\leftmargini}{1em}
\begin{itemize}
\vspace{-3mm}
    \item \emph{Correct} (binary): Indicates whether the feedback provides correct information that can help with debugging the code.
    \vspace{-1mm}
    \item \emph{Num. words} (integer): Counts the number of whitespace-separated words in the feedback.
    \vspace{-1mm}
    \item \emph{Num. sentences} (integer): Counts the number of sentences in the feedback, identified by end-of-sentence punctuation or new lines.
    \vspace{-1mm}
    \item \emph{Gives fix} (binary): Indicates whether the feedback explicitly suggests how to fix the buggy program, either conceptually or by specifying changes.
    \vspace{-1mm}
    \item \emph{Mentions variables} (binary): Indicates whether the feedback mentions specific variables from the buggy code.
    \vspace{-1mm}
    \item \emph{Mentions lines} (binary): Indicates whether the feedback mentions line numbers in the code.
\vspace{-3mm}
\end{itemize}

The attributes \emph{Num. words} and \emph{Num. sentences} were computed automatically. The remaining attributes were manually annotated, as discussed below.

\textbf{Expert annotation for student-written feedback.}
We asked two experts to evaluate the feedback in a scheme similar to prior work \cite{DBLP:conf/edm/PhungCGKMSS23,hellas2023exploring} according to the rubric described above. The experts assessed a random sample of $300$ student-written feedback instances, with $100$ instances for each of the three problems. First, each expert independently annotated $10$ instances per problem to assess inter-rater reliability. They achieved a \emph{substantial agreement} with a Cohen's kappa of $0.63$ \cite{cohen1960coefficient}.\footnote{The evaluation rubric can be refined iteratively to improve the inter-rater agreement. Future work could explore rubric refinement to further improve reliability and robustness.} Afterward, one expert annotated the entire sample of student feedback.

\textbf{Expert annotation for AI-generated feedback.}
As described in Section~\ref{sec.setup.student}, there is a set of $10$ buggy programs for each of the three problems. We generated the feedback using all of the models for each of the programs.
The same two experts mentioned above independently annotated each instance based on the described rubric.

%% file: figs/setup/study_workflow.tex
\begin{figure}
    \centering
    \includegraphics[width=0.99\linewidth]{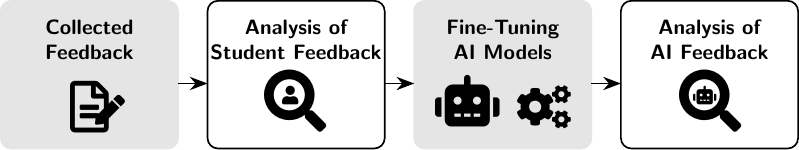}

    \Description{A diagram showing the workflow of the data analysis. The process starts from collected student-written feedback, goes through analysis of student feedback, continues with fine-tuning of AI models, and finally moves to the analysis of AI-generated feedback.}
    
    \caption{Data analysis workflow (Sections~\ref{sec.setup.technique}~and~\ref{sec.setup.rubric})}
    \label{fig.setup.workflow}      
\end{figure}

%% file: figs/finetuning/prompt_overview.tex
\begin{figure}[t!]
\Description{Examples of a basic prompt and an engineered prompt for feedback generation provided to AI model.}
    \begin{subfigure}{.98\linewidth}
        \scalebox{0.965}{
            \input{figs/finetuning/src/prompt_basic}
        }
        \caption{Basic prompt}
        \hfill
        \label{fig.prompt_basic}
    \end{subfigure}
    \begin{subfigure}{.98\linewidth}
        \scalebox{0.965}{
            \input{figs/finetuning/src/prompt}
        }
        \caption{Engineered prompt}
        \label{fig.prompt}
    \end{subfigure}
    \caption{Comparison of prompts using two kinds of instructions (\hl{highlighted in yellow}) for generating feedback.}
    \label{fig:prompts_comparison}
\end{figure}

%% file: figs/finetuning/src/prompt_basic.tex
\begin{tabular}{|p{0.98\linewidth}|}
    \hline
    \multicolumn{1}{|c|}{\color{RoyalPurple} Basic Prompt for Feedback Generation} \\ 
    \small
    I am working on a C programming problem. The current program below is not working well. Can you help by giving feedback?
    \newline
    \newline
    \begin{tabular}{@{}p{0.61\linewidth}r}
    Problem description:  &   {\color{OliveGreen}\{problem\_description\}} \\
    Failing test cases:   &   {\color{OliveGreen}\{failing\_test\_case\}} \\
    Buggy program:        &   {\color{OliveGreen}\{buggy\_program\}} \\
    Fixed program for the buggy program: & {\color{OliveGreen}\{fixed\_program\}} \\
    \end{tabular}
    \newline
    \newline 
    \hl{Can you help by giving feedback?}
    \\
    \hline
\end{tabular}

%% file: figs/finetuning/src/prompt.tex
\begin{tabular}{|p{0.98\linewidth}|}
    \hline
    \multicolumn{1}{|c|}{\color{RoyalPurple} Engineered Prompt for Feedback Generation} \\ 
    \small
    I am working on a C programming problem. The current program below is not working well. Can you help by giving feedback?
    \newline
    \newline
    \begin{tabular}{@{}p{0.61\linewidth}r}
    Problem description:  &   {\color{OliveGreen}\{problem\_description\}} \\
    Failing test cases:   &   {\color{OliveGreen}\{failing\_test\_case\}} \\
    Buggy program:        &   {\color{OliveGreen}\{buggy\_program\}} \\
    Fixed program for the buggy program: & {\color{OliveGreen}\{fixed\_program\}} \\
    \end{tabular}
    \newline
    \newline
    \hl{
    \! 1. Describe the bugs and provide an explanation along with fixes.}
    \newline  
    \hl{
    \; 2. Provide a concise hint about one bug in the buggy code. Do not give out the answer or any code. If there's an obvious bug, direct to the location of the bug. If there's a conceptual misunderstanding, offer a conceptual refresher. Limit your response for the hint to a sentence or two at most. Be as socratic as possible, and be super friendly.}
    \\
    \hline
\end{tabular}

%% file: figs/results/fig_overview.tex

\begin{table*}
\caption{Results for a sample of 300 instances of student-written feedback. We first break down the numbers per problem, then by the student's ability to provide a test case (i.e., Understanding), and finally we combine the two conditions.
}
\vspace{-2mm}
\label{fig.results}
\centering{
\scalebox{0.92}{
\setlength\tabcolsep{9.2pt}
\renewcommand{\arraystretch}{1.05}  
    \begin{tabular}{l | c | c  c  c  c  c  c }
    \toprule
    \textbf{Problem\{1/2/3\}:} &
    \textbf{Sample} &
    \multirow{2}{*}{\textbf{Correct \%}} &
    \textbf{Num.} &
    \textbf{Num.} &
    \textbf{Gives} &
    \textbf{Mentions} &
    \textbf{Mentions} \\
    \textbf{Understanding\{Any/1/0\}} &
    \textbf{Size} &
    &
    \textbf{Words} &
    \textbf{Sentences} &
    \textbf{Fix \%} &
    \textbf{Variables \%} &
    \textbf{Lines \%} \\
    \midrule
    Three problems:Understanding=Any &
    $300$ &
    $77.3$ &
    $46.1$ &
    $2.7$ &
    $46.0$ &
    $36.3$ &
    $11.3$ \\
    \midrule
    Problem 1:Understanding=Any &
    $100$ &
    $78.0$ &
    $49.4$ &
    $2.9$ &
    $48.0$ &
    $41.0$ &
    $14.0$ \\
    Problem 2:Understanding=Any &
    $100$ &
    $79.0$ &
    $47.4$ &
    $2.8$ &
    $48.0$ &
    $31.0$ &
    $\phantom{0}9.0$ \\
    Problem 3:Understanding=Any &
    $100$ &
    $75.0$ &
    $42.3$ &
    $2.6$ &
    $42.0$ &
    $37.0$ &
    $11.0$ \\
    \midrule
    Three problems:Understanding=1 &
    $172$ &
    $84.3$ &
    $49.5$ &
    $2.9$ &
    $49.4$ &
    $40.7$ &
    $14.5$ \\
    Three problems:Understanding=0 & 
    $128$ &
    $68.0$ &
    $41.5$ &
    $2.5$ &
    $41.4$ &
    $30.5$ &
    $\phantom{0}7.0$ \\
    \cmidrule{1-8}
    Problem 1:Understanding=1 &
    $\phantom{0}74$ &
    $85.1$ &
    $50.3$ &
    $2.9$ &
    $51.4$ &
    $40.5$ &
    $14.9$ \\
    Problem 1:Understanding=0 & 
    $\phantom{0}26$ &
    $57.7$ &
    $43.8$ &
    $2.7$ &
    $38.5$ &
    $42.3$ &
    $11.5$ \\
    Problem 2:Understanding=1 &
    $\phantom{0}47$ &
    $83.0$ &
    $57.2$ &
    $3.4$ &
    $53.2$ &
    $36.2$ &
    $10.6$ \\
    Problem 2:Understanding=0 & 
    $\phantom{0}53$ &
    $75.5$ &
    $38.8$ &
    $2.3$ &
    $43.4$ &
    $26.4$ &
    $\phantom{0}7.5$ \\
    Problem 3:Understanding=1 &
    $\phantom{0}51$ &
    $84.3$ &
    $41.5$ &
    $2.4$ &
    $43.1$ &
    $45.1$ &
    $17.6$ \\
    Problem 3:Understanding=0 & 
    $\phantom{0}49$ &
    $65.3$ &
    $43.2$ &
    $2.7$ &
    $40.8$ &
    $28.6$ &
    $\phantom{0}4.1$ \\
    \bottomrule
    \end{tabular}
    } 
    } 
\end{table*}

%% file: 4_results.tex

\input{figs/finetuning/results_fig_overview}

\section{Results}\label{sec.results}

In this section, we analyze the characteristics of student-written feedback and evaluate the effectiveness of fine-tuning generative models to align with it. First, we examine the correctness, conciseness, and structure of student feedback, considering factors such as problem complexity and student understanding of the bug. Then, we compare AI-generated feedback across different prompting strategies and fine-tuned models, assessing how well they replicate student feedback in both style and accuracy.

\subsection{Analysis of Student-Written Feedback} \label{sec.result.rq1}

To understand the characteristics of student-written feedback, we analyze three key aspects: (a) whether students provide correct feedback and if they prefer concise hints or detailed explanations, (b) how problem complexity influences feedback characteristics, and (c) how a student’s understanding of the bug affects the quality and detail of their feedback.

\textbf{Student feedback is mostly correct and concise.} First, we explore the student-written feedback in terms of correctness and style. As shown by the first row in Table~\ref{fig.results}, the overall feedback correctness is high, approaching $80\%$. This indicates that students are generally capable of providing accurate feedback for programming bugs. We note that this high rate could be partly affected by the fact that we tried to increase feedback correctness by showing the students the fixed program, as described in Section~\ref{sec.setup}. A feedback instance on average comprises only $46$ words spanning fewer than $3$ sentences, with slightly fewer than half of them explicitly mentioning how to fix the buggy code. This suggests that students generally favor short, concise feedback but some students are flexible and include actionable ideas for fixing the code. Additionally, we observe that students prefer using variable names rather than line numbers for localizing bugs and fixes.

\textbf{Problem complexity has little effect on feedback style.} Next, we investigate whether the complexity of the problem affects the type of feedback students provide. Despite the assumption that Problem $1$ is simpler and Problems $2$ and $3$ are more complex, no clear differences emerged between the three problems in terms of feedback correctness as well as the percentages of giving fixes and mentioning variables or line numbers. The only noticeable variation is in the length of the feedback, with the number of words and sentences tending to gradually decrease. We note that this trend might be attributed to the fixed order in which the problems were presented to the students.

\textbf{Better understanding yields more targeted feedback.} Finally, we analyze whether the students' performance in understanding bugs affects their feedback style. The last eight rows in Table~\ref{fig.results} illustrate the results concerning students' understanding of the problem and buggy code, as indicated by their ability to provide a failing test case. Correctness is higher among students who provided a good test case for the bug, even though an example of a good test case and fixed code were provided afterward before they wrote feedback. Additionally, these students tend to offer longer and more detailed explanations, include fixes more frequently, and provide more targeted feedback by referencing specific aspects of the buggy code.

\subsection{Fine-Tuning for Feedback Generation}

We now analyze the impact of prompt engineering and fine-tuning on AI-generated feedback. First, we examine how prompting strategies influence the style and correctness of model-generated feedback. Then, we evaluate whether fine-tuning models on learnersourced data improves accuracy and alignment with student-written feedback.

\textbf{Engineered prompts reduce verbosity.} We first investigate the effects of prompting strategies on generated feedback, and how well the engineered prompt guides the model's output toward resembling student feedback. The first nine rows in Table~\ref{fig.finetuning.results} present a comparison between feedback from students and various generative models, differentiating on base models and prompting strategies.
It is evident that all base models using the basic prompt tend to be verbose and usually provide fixes. However, the engineered prompt substantially reduces verbosity and enhances correctness, more closely aligning the feedback style with that of students.
Figures~\ref{fig.finetuning.examples_llama.base} and ~\ref{fig.finetuning.examples_llama.prompting} show contrastive examples of the verbosity between the basic and engineered prompts.
Because the engineered prompt is based on existing expert-crafted designs, it explicitly instructs AI models not to provide fixes. While this enables direct comparison with prior work, it may not always align with how students naturally write feedback.

\textbf{Fine-tuning improves alignment and accuracy.} Next, we answer whether fine-tuning open-access models with student-written feedback can potentially replace more complex prompt engineering approaches for generating feedback. We investigated the effectiveness of fine-tuning Llama3-8B and Phi3-3.8B. The last two rows in Table~\ref{fig.finetuning.results} indicate that both models produced short feedback, more closely resembling the style of students. The feedback generated by Llama3-8B-fine-tuned is closest to student-written feedback in terms of length and frequency of mentioning variables.
Another observation is that these AI models tend to use line numbers less frequently than students, preferring to name variables instead.
Remarkably, the correctness of AI-generated feedback improves after fine-tuning, surpassing that of the base model with the engineered prompt even though some of the feedback instances used for fine-tuning are incorrect. This improvement may result from the models becoming more familiar with the domain and generating shorter output, reducing the chances of errors.

%% file: figs/finetuning/results_fig_overview.tex

\begin{table*}
\caption{Results for the AI-generated feedback aggregated over two experts, compared to student-written feedback. We assessed the models for $10$ buggy programs for each of the $3$ problems. Rows are grouped by model family.}
\label{fig.finetuning.results}
\vspace{-2mm}
\centering{
\scalebox{0.92}{
\setlength\tabcolsep{4.6pt}
\renewcommand{\arraystretch}{0.98}
    \begin{tabular}{ll | c  c | r  r  r  r  r  r }
    \toprule 
    \multirow{2}{*}{\textbf{Feedback Source}} &
    \textbf{Prompting} &
    \multicolumn{1}{c}{\textbf{Num.}} &
    \multicolumn{1}{c|}{\textbf{Num. Programs}} &
    \multicolumn{1}{c}{\multirow{2}{*}{\textbf{Correct \%}}} &
    \multicolumn{1}{c}{\textbf{Num.}} &
    \multicolumn{1}{c}{\textbf{Num.}} &
    \multicolumn{1}{c}{\textbf{Gives}} &
    \multicolumn{1}{c}{\textbf{Mentions}} &
    \multicolumn{1}{c}{\textbf{Mentions}} \\
    &
    \textbf{Strategy} &
    \textbf{Problems} &
    \textbf{per Problem} &
    &
    \textbf{Words} &
    \textbf{Sentences} &
    \multicolumn{1}{c}{\textbf{Fix \%}} &
    \multicolumn{1}{c}{\textbf{Variables \%}} &
    \multicolumn{1}{c}{\textbf{Lines \%}} \\
    \midrule
    \multicolumn{2}{l|}{Human students} &
    $3$ &
    $100$ &
    $77.3$ &
    $46.1$ &
    $2.7\phantom{0}$ &
    $46.0$ &
    $36.3$ &
    $11.3$ \\
    \midrule
    
    GPT-4 Turbo &
    basic &
    $3$ &
    \phantom{$0$}$10$ &
    $81.7$ &
    $411.3$ &
    $47.7\phantom{0}$ &
    $96.7$ &
    $100.0$ &
    $1.7$ \\
    GPT-4 Turbo &
    engineered &
    $3$ &
    \phantom{$0$}$10$ &
    $96.7$ &
    $35.6$ &
    $2.4\phantom{0}$ &
    $6.7$ &
    $23.3$ &
    $0.0$ \\
    \midrule
    GPT-3.5 Turbo &
    basic &
    $3$ &
    \phantom{$0$}$10$ &
    $63.3$ &
    $140.2$ &
    $11.8\phantom{0}$ &
    $80.0$ &
    $71.7$ &
    $3.3$ \\
    GPT-3.5 Turbo &
    engineered &
    $3$ &
    \phantom{$0$}$10$ &
    $90.0$ &
    $20.1$ &
    $1.4\phantom{0}$ &
    $0.0$ &
    $15.0$ &
    $1.7$ \\
    \midrule
    Llama3-8B &
    basic &
    $3$ &
    \phantom{$0$}$10$ &
    $56.7$ &
    $256.3$ &
    $32.2\phantom{0}$ &
    $100.0$ &
    $90.0$ &
    $3.3$ \\
    Llama3-8B &
    engineered &
    $3$ &
    \phantom{$0$}$10$ &
    $71.7$ &
    $27.8$ &
    $3.3\phantom{0}$ &
    $3.3$ &
    $20.0$ &
    $0.0$ \\
    \midrule
    Phi3-3.8B &
    basic &
    $3$ &
    \phantom{$0$}$10$ &
    $55.0$ &
    $210.6$ &
    $35.1\phantom{0}$ &
    $100.0$ &
    $58.3$ &
    $0.0$ \\
    Phi3-3.8B &
    engineered &
    $3$ &
    \phantom{$0$}$10$ &
    $80.0$ &
    $25.7$ &
    $1.9\phantom{0}$ &
    $5.0$ &
    $5.0$ &
    $0.0$ \\
    \midrule
    Llama3-8B-fine-tuned &
    basic &
    $3$ &
    \phantom{$0$}$10$ &
    $86.7$ &
    $47.7$ &
    $2.7\phantom{0}$ &
    $71.7$ &
    $40.0$ &
    $1.7$ \\
    Phi3-3.8B-fine-tuned &
    basic &
    $3$ &
    \phantom{$0$}$10$ &
    $88.3$ &
    $68.0$ &
    $3.9\phantom{0}$ &
    $98.3$ &
    $60.0$ &
    $1.7$ \\
    \bottomrule
    \end{tabular}
    } 
    } 
\end{table*}

%% file: 5_conclusion.tex
\section{Concluding Discussions}\label{sec.conclusion}

This study explored how learnersourced data can enhance AI-generated feedback, aligning it more closely with student communication styles. We first characterized student-written feedback on buggy code, focusing on correctness, length, inclusion of explicit fixes, and methods of bug identification. Our findings reveal that students tend to write concise feedback while remaining flexible in suggesting fixes, highlighting key patterns that AI-generated feedback should replicate. Building on these insights, we fine-tuned generative models on learnersourced data, reducing reliance on expert prompt engineering. This approach improves the adaptability of AI-generated feedback, making it more peer-like and responsive to diverse student needs. Notably, even when some student feedback was incorrect, fine-tuning still led to substantial improvements in accuracy, showcasing the potential of student-centered fine-tuning for scalable and personalized feedback systems in programming education.

While our findings are promising, several limitations suggest avenues for future research. First, we did not evaluate the models' effectiveness in real classroom settings. Deploying these models in real classrooms and gathering student reflections on their utility and relevance would provide insights into their practical impact. Second, students in our study were given a test case and a corrected program to maximize feedback accuracy. While this ensured high correctness rates (close to $80\%$), it may not reflect real-world scenarios where such guidance is unavailable. Future work could examine how feedback quality and correctness change when students must identify bugs without prior guidance, evaluating model robustness in more realistic settings. Third, we did not explore the potential of fine-tuning models for specific courses, which could improve feedback relevance and better support course-specific learning needs. Employing learnersourced data in this context also introduces ethical considerations, such as ensuring data protection, respecting intellectual property rights, and transparently communicating how content created by students is utilized. Although our analysis did not reveal that models inherited misconceptions from student-written feedback, future applications should explicitly monitor and mitigate this risk. Addressing these ethical and methodological aspects thoughtfully can strengthen the reliability and effectiveness of fine-tuned generative models. Ultimately, combining crowdsourcing with fine-tuning remains a promising and scalable approach to delivering personalized, high-quality feedback, especially valuable in low-resource educational settings.

%% file: 6_acknowledgements.tex
\section*{Acknowledgments}
Juho Leinonen acknowledges funding by Research Council of Finland (Academy Research Fellow grant number 356114). Michael Liut acknowledges funding by NSERC Discovery Grant \#RGPIN-2024-04348. Funded/Co-funded by the European Union (ERC, TOPS, 101039090). Views and opinions expressed are however those of the author(s) only and do not necessarily reflect those of the European Union or the European Research Council. Neither the European Union nor the granting authority can be held responsible for them.

%% file: main.bbl
\begin{thebibliography}{10}

\bibitem{DBLP:journals/corr/abs-2404-14219}
M.~I. Abdin et~al.
\newblock {P}hi-3 {T}echnical {R}eport: {A} {H}ighly {C}apable {L}anguage {M}odel {L}ocally on {Y}our {P}hone.
\newblock {\em CoRR}, abs/2404.14219, 2024.

\bibitem{al2018review}
A.~Al-batlaa, M.~Abdullah-Al-Wadud, and M.~A. Hossain.
\newblock {A} {R}eview on {R}ecommending {S}olutions for {B}ugs {U}sing {C}rowdsourcing.
\newblock In {\em Saudi Computer Society National Computer Conference {(NCC)}}, 2018.

\bibitem{ala2005survey}
K.~M. Ala-Mutka.
\newblock A {S}urvey of {A}utomated {A}ssessment {A}pproaches for {P}rogramming {A}ssignments.
\newblock {\em Computer Science Education}, 2005.

\bibitem{becker2016effective}
B.~A. Becker.
\newblock {A}n {E}ffective {A}pproach to {E}nhancing {C}ompiler {E}rror {M}essages.
\newblock In {\em Proc. of the Technical Symp. on Computer Science Education (SIGCSE)}, 2016.

\bibitem{cohen1960coefficient}
J.~Cohen.
\newblock {A} {C}oefficient of {A}greement for {N}ominal {S}cales.
\newblock {\em Educationalf and Psychological Measurement}, 1960.

\bibitem{DBLP:journals/corr/abs-2402-01580}
P.~Denny, S.~Gulwani, N.~T. Heffernan, T.~K{\"a}ser, S.~Moore, A.~N. Rafferty, and A.~Singla.
\newblock {G}enerative {AI} for {E}ducation ({GAIED}): {A}dvances, {O}pportunities, and {C}hallenges.
\newblock {\em CoRR}, abs/2402.01580, 2024.

\bibitem{frazier2024customizing}
M.~Frazier, K.~Damevski, and L.~Pollock.
\newblock {C}ustomizing {C}hat{GPT} to {H}elp {C}omputer {S}cience {P}rinciples {S}tudents {L}earn {T}hrough {C}onversation.
\newblock In {\em Proceedings of the Conference on Innovation and Technology in Computer Science Education {(ITiCSE)}}, 2024.

\bibitem{hartmann2010would}
B.~Hartmann, D.~MacDougall, J.~Brandt, and S.~R. Klemmer.
\newblock {W}hat {W}ould {O}ther {P}rogrammers {D}o: {S}uggesting {S}olutions to {E}rror {M}essages.
\newblock In {\em Proceedings of the International Conference on Human Factors in Computing Systems {(CHI)}}, 2010.

\bibitem{head2017writing}
A.~Head, E.~L. Glassman, G.~Soares, R.~Suzuki, L.~Figueredo, L.~D'Antoni, and B.~Hartmann.
\newblock {W}riting {R}eusable {C}ode {F}eedback at {S}cale with {M}ixed-{I}nitiative {P}rogram {S}ynthesis.
\newblock In {\em Proceedings of the Conference on Learning @ Scale (L@S)}, 2017.

\bibitem{hellas2023exploring}
A.~Hellas, J.~Leinonen, S.~Sarsa, C.~Koutcheme, L.~Kujanp{\"a}{\"a}, and J.~Sorva.
\newblock {E}xploring the {R}esponses of {L}arge {L}anguage {M}odels to {B}eginner {P}rogrammers’ {H}elp {R}equests.
\newblock In {\em Proceedings of the Conference on International Computing Education Research {(ICER)}}, 2023.

\bibitem{ihantola2010review}
P.~Ihantola, T.~Ahoniemi, V.~Karavirta, and O.~Sepp{\"a}l{\"a}.
\newblock {R}eview of {R}ecent {S}ystems for {A}utomatic {A}ssessment of {P}rogramming {A}ssignments.
\newblock In {\em Proceedings of the Koli Calling International Conference on Computing Education Research}, 2010.

\bibitem{islam2024mapcoder}
M.~A. Islam et~al.
\newblock {M}ap{C}oder: {M}ulti-{A}gent {C}ode {G}eneration for {C}ompetitive {P}roblem {S}olving.
\newblock In {\em Proceedings of the Annual Meeting of the Association for Computational Linguistics ({ACL})}, 2024.

\bibitem{jiang2018toward}
H.~Jiang et~al.
\newblock {T}oward {B}etter {S}ummarizing {B}ug {R}eports with {C}rowdsourcing {E}licited {A}ttributes.
\newblock {\em IEEE Transactions on Reliability}, 2018.

\bibitem{khosravi2023learnersourcing}
H.~Khosravi, P.~Denny, S.~Moore, and J.~Stamper.
\newblock {L}earnersourcing in the {A}ge of {AI}: {S}tudent, {E}ducator and {M}achine {P}artnerships for {C}ontent {C}reation.
\newblock {\em Computers and Education: Artificial Intelligence}, 2023.

\bibitem{kohn2020tell}
T.~Kohn and B.~Manaris.
\newblock {T}ell {M}e {W}hat's {W}rong: {A} {P}ython {IDE} with {E}rror {M}essages.
\newblock In {\em Proceedings of the Technical Symposium on Computer Science Education (SIGCSE)}, 2020.

\bibitem{kotalwar2024hintsinbrowser}
N.~Kotalwar, A.~Gotovos, and A.~Singla.
\newblock {Hints-In-Browser}: {B}enchmarking {L}anguage models for {P}rogramming {F}eedback {G}eneration.
\newblock In {\em Proceedings of the Annual Conference on Neural Information Processing Systems ({NeurIPS}) Track on Datasets and Benchmarks}, 2024.

\bibitem{koutcheme2024open}
C.~Koutcheme, N.~Dainese, S.~Sarsa, A.~Hellas, J.~Leinonen, and P.~Denny.
\newblock Open source language models can provide feedback: Evaluating llms' ability to help students using gpt-4-as-a-judge.
\newblock In {\em Proceedings of the Conference on Innovation and Technology in Computer Science Education {(ITiCSE)}}, 2024.

\bibitem{DBLP:journals/corr/abs-2310-13712}
H.~Kumar, I.~Musabirov, M.~Reza, J.~Shi, A.~Kuzminykh, J.~J. Williams, and M.~Liut.
\newblock {I}mpact of {G}uidance and {I}nteraction {S}trategies for {LLM} {U}se on {L}earner {P}erformance and {P}erception.
\newblock {\em CoRR}, abs/2310.13712, 2023.

\bibitem{lahza2023analytics}
H.~Lahza et~al.
\newblock {A}nalytics of {L}earning {T}actics and {S}trategies in an {O}nline {L}earnersourcing {E}nvironment.
\newblock {\em Journal of Computer Assisted Learning}, 2023.

\bibitem{DBLP:conf/sigcse/0001HSRDPB23}
J.~Leinonen, A.~Hellas, S.~Sarsa, B.~N. Reeves, P.~Denny, J.~Prather, and B.~A. Becker.
\newblock {U}sing {L}arge {L}anguage {M}odels to {E}nhance {P}rogramming {E}rror {M}essages.
\newblock In {\em Proceedings of the Technical Symposium on Computer Science Education (SIGCSE)}, 2023.

\bibitem{DBLP:conf/sigcse/LiuZLHTM24}
R.~Liu, C.~Zenke, C.~Liu, A.~Holmes, P.~Thornton, and D.~J. Malan.
\newblock {T}eaching {CS50} with {AI:} {L}everaging {G}enerative {A}rtificial {I}ntelligence in {C}omputer {S}cience {E}ducation.
\newblock In {\em Proceedings of the Technical Symposium on Computer Science Education (SIGCSE)}, 2024.

\bibitem{DBLP:conf/iticse/LiuYHBBL24}
S.~Liu, Z.~Yu, F.~Huang, Y.~Bulbulia, A.~Bergen, and M.~Liut.
\newblock {C}an {S}mall {L}anguage {M}odels {W}ith {R}etrieval-{A}ugmented {G}eneration {R}eplace {L}arge {L}anguage {M}odels {W}hen {L}earning {C}omputer {S}cience?
\newblock In {\em Proceedings of Innovation and Technology in Computer Science Education (ITiCSE)}, 2024.

\bibitem{Llama-3}
Meta.
\newblock Llama-3.
\newblock \url{https://ai.meta.com/blog/meta-llama-3/}, 2024.

\bibitem{mujumdar2011crowdsourcing}
D.~Mujumdar, M.~Kallenbach, B.~Liu, and B.~Hartmann.
\newblock {C}rowdsourcing {S}uggestions to {P}rogramming {P}roblems for {D}ynamic {W}eb {D}evelopment {L}anguages.
\newblock In {\em Extended Abstracts on Human Factors in Computing Systems {(CHI)}}, 2011.

\bibitem{ChatGPT}
OpenAI.
\newblock Chat{GPT}.
\newblock \url{https://openai.com/blog/chatgpt}, 2023.

\bibitem{GPT4}
OpenAI.
\newblock {GPT-4} {T}echnical {R}eport.
\newblock {\em CoRR}, abs/2303.08774, 2023.

\bibitem{GPT-Family}
OpenAI.
\newblock {O}pen{AI} {P}latform {M}odels.
\newblock \url{https://platform.openai.com/docs/models}, 2024.

\bibitem{otaki2024generative}
B.~Otaki and O.~Lindwall.
\newblock {G}enerative {AI} and the {H}uman {T}ouch: {I}nvestigating the {C}hanging {L}andscape of {F}eedback in {H}igher {E}ducation.
\newblock In {\em Proceedings of the International Conference of the Learning Sciences (ICLS)}, 2024.

\bibitem{DBLP:conf/sigcse/Padurean0S25}
V.~Padurean, P.~Denny, and A.~Singla.
\newblock {B}ug{S}potter: {A}utomated {G}eneration of {C}ode {D}ebugging {E}xercises.
\newblock In {\em Proceedings of the Technical Symposium on Computer Science Education (SIGCSE)}, 2025.

\bibitem{paiva2022automated}
J.~Paiva et~al.
\newblock {A}utomated {A}ssessment in {C}omputer {S}cience {E}ducation: A {S}tate-of-the-{A}rt {R}eview.
\newblock {\em Transactions on Computing Education ({TOCE})}, 2022.

\bibitem{DBLP:conf/edm/PhungCGKMSS23}
T.~Phung, J.~Cambronero, S.~Gulwani, T.~Kohn, R.~Majumdar, A.~Singla, and G.~Soares.
\newblock {G}enerating {H}igh-{P}recision {F}eedback for {P}rogramming {S}yntax {E}rrors using {L}arge {L}anguage {M}odels.
\newblock In {\em Proceedings of the International Conference on Educational Data Mining (EDM)}, 2023.

\bibitem{DBLP:journals/corr/abs-2306-17156}
T.~Phung, V.~P{\u a}durean, J.~Cambronero, S.~Gulwani, T.~Kohn, R.~Majumdar, A.~Singla, and G.~Soares.
\newblock {G}enerative {AI} for {P}rogramming {E}ducation: {B}enchmarking {C}hatgpt, {GPT}-4, and {H}uman {T}utors.
\newblock In {\em Proceedings of the Conference on International Computing Education Research (ICER) - Volume 2}, 2023.

\bibitem{DBLP:conf/lak/PhungPS0CGSS24}
T.~Phung, V.~Padurean, A.~Singh, C.~Brooks, J.~Cambronero, S.~Gulwani, A.~Singla, and G.~Soares.
\newblock {A}utomating {H}uman {T}utor-{S}tyle {P}rogramming {F}eedback: {L}everaging {GPT}-4 {T}utor {M}odel for {H}int {G}eneration and {GPT}-3.5 {S}tudent {M}odel for {H}int {V}alidation.
\newblock In {\em Proceedings of the International Learning Analytics and Knowledge Conference (LAK)}, 2024.

\bibitem{piech2015learning}
C.~Piech, J.~Huang, A.~Nguyen, M.~Phulsuksombati, M.~Sahami, and L.~J. Guibas.
\newblock {L}earning {P}rogram {E}mbeddings to {P}ropagate {F}eedback on {S}tudent {C}ode.
\newblock In {\em Proceedings of the International Conference on Machine Learning (ICML)}, 2015.

\bibitem{pirttinen2023lessons}
N.~Pirttinen, P.~Denny, A.~Hellas, and J.~Leinonen.
\newblock {L}essons {L}earned from {F}our {C}omputing {E}ducation {C}rowdsourcing {S}ystems.
\newblock {\em IEEE Access}, 2023.

\bibitem{pirttinen2022can}
N.~Pirttinen and J.~Leinonen.
\newblock {C}an {S}tudents {R}eview {T}heir {P}eers? {C}omparison of {P}eer and {I}nstructor {R}eviews.
\newblock In {\em Proceedings of the Annual Conference on Innovation and Technology in Computer Science Education (ITiCSE)}, 2022.

\bibitem{singh2022learnersourcing}
A.~Singh, C.~Brooks, and S.~Doroudi.
\newblock {L}earnersourcing in {T}heory and {P}ractice: {S}ynthesizing the {L}iterature and {C}harting the {F}uture.
\newblock In {\em Proceedings of the Conference on Learning @ Scale (L@S)}, 2022.

\bibitem{singh2021s}
A.~Singh, C.~Brooks, Y.~Lin, and W.~Li.
\newblock {W}hat's in {I}t for the {L}earners? {E}vidence from a {R}andomized {F}ield {E}xperiment on {L}earnersourcing {Q}uestions in a {MOOC}.
\newblock In {\em Proceedings of the Conference on Learning @ Scale (L@S)}, 2021.

\bibitem{singh2024bridging}
A.~Singh, C.~Brooks, X.~Wang, W.~Li, J.~Kim, and D.~Wilson.
\newblock {B}ridging {L}earnersourcing and {AI}: {E}xploring the {D}ynamics of {S}tudent-{AI} {C}ollaborative {F}eedback {G}eneration.
\newblock In {\em Proceedings of the International Learning Analytics and Knowledge Conference (LAK)}, 2024.

\bibitem{singh2013automated}
R.~Singh et~al.
\newblock {A}utomated {F}eedback {G}eneration for {I}ntroductory {P}rogramming {A}ssignments.
\newblock In {\em Proceedings of the Conference on Programming Language Design and Implementation {(PLDI)}}, 2013.

\bibitem{DBLP:conf/sigcse/WangMP24}
S.~Wang, J.~C. Mitchell, and C.~Piech.
\newblock {A} {L}arge {S}cale {RCT} on {E}ffective {E}rror {M}essages in {CS1}.
\newblock In {\em Proceedings of the Technical Symposium on Computer Science Education (SIGCSE)}, 2024.

\bibitem{qi2023conversational}
J.~Zamfirescu-Pereira, L.~Qi, B.~Hartmann, J.~DeNero, and N.~Norouzi.
\newblock {C}onversational {P}rogramming with {LLM}-{P}owered {I}nteractive {S}upport in an {I}ntroductory {C}omputer {S}cience {C}ourse.
\newblock {\em NeurIPS'23 Workshop on Generative AI for Education (GAIED)}, 2023.

\bibitem{DBLP:journals/corr/abs-2403-13372}
Y.~Zheng, R.~Zhang, J.~Zhang, Y.~Ye, Z.~Luo, and Y.~Ma.
\newblock {L}lama{F}actory: {U}nified {E}fficient {F}ine-{T}uning of 100+ {L}anguage {M}odels.
\newblock In {\em Proceedings of the Annual Meeting of the Association for Computational Linguistics ({ACL})}, 2024.

\bibitem{DBLP:conf/nips/ZhouLX0SMMEYYZG23}
C.~Zhou et~al.
\newblock {LIMA:} {L}ess {I}s {M}ore for {A}lignment.
\newblock In {\em Proceedings of the Annual Conference on Neural Information Processing Systems ({NeurIPS})}, 2023.

\end{thebibliography}
